\documentclass[lettersize,journal]{IEEEtran}
\usepackage{amsmath,amsfonts}
\usepackage{algorithmic}
\usepackage{algorithm}
\usepackage{array}
\usepackage[caption=false,font=normalsize,labelfont=sf,textfont=sf]{subfig}
\usepackage{textcomp}
\usepackage{stfloats}
\usepackage{hyperref}
\usepackage{verbatim}
\usepackage{graphicx}
\usepackage{xcolor}
\usepackage{cite}
\usepackage{tikz}
\usetikzlibrary{positioning,fit,arrows.meta,calc}

\tikzset{
block/.style={
draw,
thick,
rectangle,
rounded corners=2pt,
minimum width=2cm,
minimum height=1cm,
align=center,
font=\small
},
dsp/.style={
draw,
thick,
rectangle,
minimum width=2.5cm,
minimum height=0.7cm,
text width=3.5cm,
align=center,
font=\small
},
line/.style={
-{Latex[length=2mm]},
thick
},
circ/.style={
draw,
circle,
minimum size=1cm,
thick
}
}

\begin{document}

\title{Distributed Acoustic Sensing Over Deployed Telecommunication Fiber Networks: Challenges, Opportunities and Recent Advances}

\author{Élie Awwad, Maria Freire Hermelo, Diane Prato, Pierre Pruvost, Pallab Kumar Choudhury, Yves Jaouën, Renaud Gabet, and Heming Huang
\thanks{Manuscript created in July 2026. All authors are with LTCI Lab, Télécom Paris, Institut Polytechnique de Paris, 19 place Marguerite Perey, 91120 Palaiseau, France.  (Corresponding author: elie.awwad@telecom-paris.fr)}}



\maketitle

\begin{abstract}
This paper discusses key challenges and opportunities associated with deploying distributed acoustic sensing (DAS) over existing optical telecommunication networks. We review recent approaches to DAS interrogation and their compatibility with data transmission over deployed fibers, and present complementary investigations addressing three challenges: the influence of fiber-to-environment mechanical coupling, the single-shot extraction of distributed birefringence information using polarization-diversity DAS, and the impact of laser phase noise on continuous-wave phase-encoded DAS. 
These investigations rely on different forms of evidence: field measurements on a deployed telecommunication cable, laboratory experiments using fiber spools, and numerical simulations, respectively. The results illustrate complementary aspects of DAS deployment, from the interpretation of measured strain to polarization-sensitive sensing and phase-noise mitigation. They also highlight the need for further validation and integration of these approaches under operational telecommunication-network conditions.
\end{abstract}


\section{Introduction}
\IEEEPARstart{A}{s} human environments and their supporting infrastructures become increasingly dense and interconnected, ensuring their safety, integrity and sustainability has become a major societal concern. The breadth and heterogeneity of these environments make their monitoring a  technological challenge. The early detection of hazards, accidental failures, or malicious damage affecting critical infrastructures is essential, as such events can have significant consequences for both operators and end users. Today, infrastructure monitoring relies on a wide range of discrete sensors, including surveillance cameras, personal devices, magnetometers, radar-based sensors, and handheld inspection equipment, to monitor road traffic~\cite{Lindsey20}, detect intrusions, identify anomalies such as road degradation~\cite{Li25}, water distribution leaks~\cite{Sharif25}, or even identify the layout of the telecommunication cable itself~\cite{Tovar26}. While these sensing modalities are effective for localized monitoring, they inherently suffer from limited spatial coverage, and may also suffer from reduced sensitivity and non-negligible susceptibility to electromagnetic interference. Furthermore, extending their deployment over large-scale infrastructures is often impractical due to the associated installation, operation, and maintenance costs. In contrast, telecommunication optical fiber networks already span cities and interconnect urban and rural areas over thousands of kilometers, making them an attractive sensing infrastructure whose reuse could enable continuous wide-area monitoring without requiring the deployment of additional sensors. 

Several distributed optical fiber sensing (DOFS) technologies have been developed, each relying on a different light-scattering mechanism. Raman-based sensing is primarily dedicated to distributed temperature measurements~\cite{Dakin85}, whereas Brillouin-based techniques provide absolute temperature and strain measurements along the fiber, albeit with limited mechanical bandwidth, which requires additional techniques to decouple temperature and strain effects~\cite{Mohamad14,Alahbabi05,Clement21}. Rayleigh-based sensing, commonly known as distributed acoustic sensing (DAS), instead measures relative variations of the optical phase using a highly coherent laser source. Although it does not directly provide absolute physical quantities, DAS offers significantly higher sensitivity and broader mechanical bandwidth, making it particularly well suited for vibration monitoring over existing telecommunication fiber infrastructures~\cite{Cedilnik19,Lindsey20,Zhu21,AjoFranklin19,Huang20,Guerrier22}. In this paper, we cover challenges and opportunities of using DAS systems over telecommunication fiber networks to monitor relative strain variations.

Despite its strong potential, deploying DAS over lit telecommunication networks remains challenging. Conventional $\Delta\phi$-OTDR DAS systems are fundamentally limited by the weak Rayleigh backscattered signal and laser phase noise. Dealing with the large volume of sensing data that must be acquired and processed is an extra challenge. Additional constraints arise when moving from dedicated sensing fibers to deployed telecommunication infrastructures. On lit fibers, sensing must coexist with communication channels without degrading network performance, while, on both lit and dark fibers, the sensing system must operate over complex network topologies including point-to-multipoint fiber links encountered in passive optical networks (PON)~\cite{Takahashi25,Choudhury25}. Recent demonstrations have highlighted the ability of DAS to monitor road traffic, human activity, and seismic events over deployed networks~\cite{Lindsey20,AjoFranklin19,Guerrier22,Huang23,Westbrook23,Mazur25}. However, these demonstrations generally involve compromises in terms of acoustic bandwidth, sensitivity, sensing range, or deployment complexity. In several cases, coexistence with communication signals required sacrificing valuable optical bandwidth~\cite{Huang20,Huang23}, while others relied on dedicated or non-standard fiber infrastructures~\cite{Cedilnik19,Westbrook23}.

Addressing these limitations requires the development of DAS architectures that can seamlessly integrate into existing \newpage \noindent telecommunication networks while preserving their operational constraints. Such systems should ultimately minimize additional infrastructure and equipment requirements to enable cost-effective and low-footprint deployment, exploit richer signal information to improve sensing sensitivity and coverage, mitigate the impact of noise through advanced signal processing and estimation techniques, and efficiently process the massive data streams generated by distributed sensing. 

This paper addresses complementary challenges associated with DAS deployment over telecommunication fiber networks and presents representative investigations that illustrate potential approaches to addressing them. These investigations are not intended to constitute a single integrated demonstration under a common deployment scenario. Instead, the paper combines a review of interrogation approaches for coexistence with data transmission, field measurements addressing fiber-to-environment coupling, laboratory experiments investigating polarization-related sensing capabilities, and numerical simulations assessing laser phase-noise effects. This complementary approach allows us to examine several important aspects of DAS deployment while distinguishing experimentally demonstrated results from potential solutions and open research questions.

The remainder of this paper is organized as follows. Section~\ref{sec:Coexist} reviews DAS interrogators and focuses on continuous-wave DAS interrogators for in-band sensing and communication coexistence. Section~\ref{sec:Coupling} investigates the effect of fiber mechanical coupling on distributed strain measurements. Section~\ref{sec:Birefringence} presents the use of polarization diversity for extracting distributed birefringence variations. Section~\ref{sec:PhaseNoise} analyzes the impact of laser phase noise on phase-encoded continuous-wave DAS interrogation. Finally, Section~\ref{sec:Conclusion} concludes the paper. 

\section{Coexistence With Communication and Adaptation to Network Topology}\label{sec:Coexist}

\begin{table*}[t]
\centering
\begin{tabular}{ >{\raggedright\arraybackslash}p{2cm} >{\raggedright\arraybackslash}p{3.5cm} >{\raggedright\arraybackslash}p{3.5cm}  >{\raggedright\arraybackslash}p{3.5cm}  >{\raggedright\arraybackslash}p{4cm} } 
  \textbf{DAS approach} & \textbf{Probing \& processing} & \textbf{Main advantages} & \textbf{Main limitations} & \textbf{Main results}\\ 
  \hline
  Conventional $\Delta\phi$-OTDR & Short optical pulses; coherent detection; phase estimated from time-resolved Rayleigh backscatter; differential phase evaluated over a gauge length  & Direct time-of-flight localization; established distributed phase sensing  & Trade-off among spatial resolution (SR), sensitivity, sensing range, and refresh rate; high peak power induces nonlinear interference (NLI)~\cite{Wang25}; phase unwrapping limits dynamic range & Used as the reference approach in the comparisons discussed in~\cite{Pastor16,Ip22,Wang25}\\ 
  \hline
  Chirped-pulse $\phi$-OTDR & Chirped optical pulses; direct detection; time-delay-based measurement  & Addresses the dynamic-range limitations associated with conventional $\Delta\phi$-OTDR & Performance remains dependent on pulse design; inherits limitations of conv. scheme except dynamic range & \cite{Pastor16} and a recent adaptation with coherent detection in~\cite{Orsuti25} \\ 
  \hline
  LFM-pulse-based DAS & Linear-frequency-modulated pulses; coherent detection; digital matched filtering  & Sensitivity gains compared to conv. scheme; SR determined by sweeping range  & Performance depends on waveform and receiver design; inherits limitations of conv. scheme & SSMF without inline amplifiers: $5$~km range, SR=$0.5$~m\cite{Zou15}, $40$~km, SR=$3.5$~m\cite{Wang17}, $10$~km, SR=$10$~m\cite{Waagaard21}; SSMF with inline amp. $1000$~km, SR=$20$~m\cite{Ip22}\\ 
  \hline
Continuous-wave (CW) phase-encoded DAS & Continuous constant-power probe using coded sequences; coherent detection; digital matched filtering & Reduced NLI relative to high-peak-power pulsed probing; sensitivity gains; suited to in-band coexistence &  Requires suitable phase-encoding and more components; inherits the other limitations of conv. scheme & SSMF without inline amplifiers: $82$~km range, SR=$100$~m\cite{Guerrier22} \\ 
  \hline
  OFDR & CW frequency-swept laser; coherent detection; matched filtering in frequency domain  & Reduced NLI; high spatial resolution; sensitivity gains; close to CW phase-encoded DAS (constant-amplitude \& matched filtering) & Requires frequency-swept interrogation and more components & SSMF without inline amplifiers: $64$~km range, SR=$6.5$~m\cite{Shiloh17}, SSMF with inline amp.:  $160$~km range, SR=$50$~m\cite{Ryf26}\\ 
  \hline
  Coherent polarization-diverse MIMO DAS & Dual-polarization probing; coherent detection; enables estimation of distributed polarization parameters & Mitigates polarization fading and provides additional sensing observables, including distributed birefringence & Requires polarization-diverse components; additional signal processing & Concept introduced in\cite{Guerrier20}; single-shot birefringence estimation in~\cite{Prato26} \& Section IV\\ 
  \hline
\end{tabular}
\vspace{1pt}
\caption{Comparison of DAS interrogation approaches and their relevance to deployed telecommunication networks \MakeLowercase{(Non-exhaustive list; Several evolutions are found in the literature handling coherent fading, phase unwrapping, or extension of refresh rate)}.}
\label{table:1}
\end{table*}

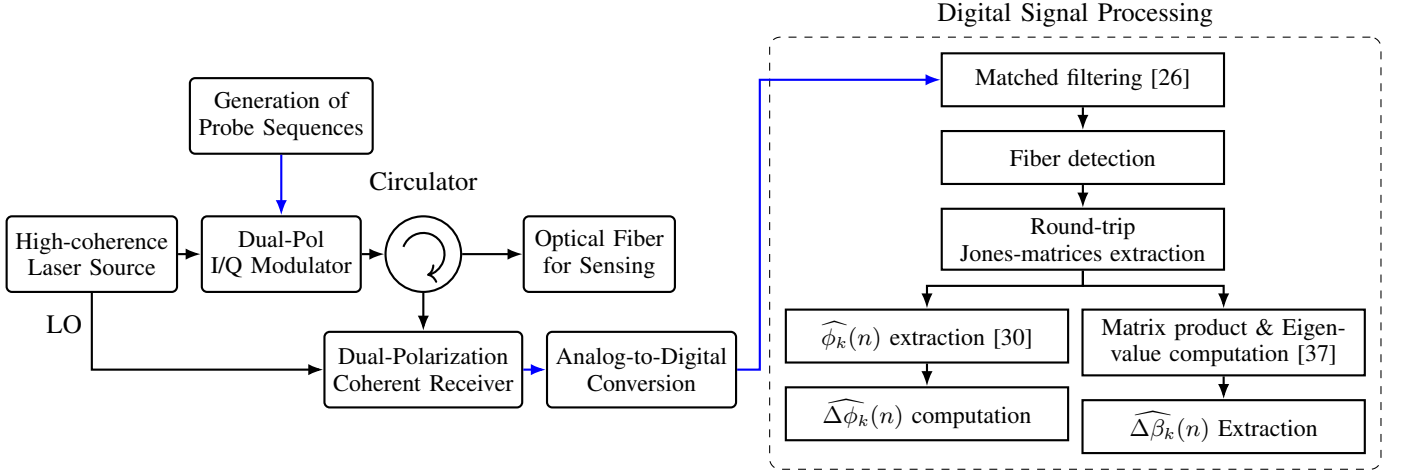
\begin{figure*}
\centering
\begin{tikzpicture}[node distance=1.5cm]

\node[block] (laser)
{High-coherence\\Laser Source};
\node[block,right=0.3cm of laser] (tx)
{Dual-Pol\\I/Q Modulator};
\node[block,above=0.8cm of tx] (seq)
{Generation of\\Probe Sequences};

\node[circ,right=0.3cm of tx] (circ){};
\draw[->,thick]
($(circ.center)+(160:0.28)$)
arc(160:-80:0.28);
\node[above=2mm of circ] {Circulator};

\node[block,right=0.8cm of circ] (fiber)
{Optical Fiber\\for Sensing};

\node[block,below=0.5cm of circ] (rx){Dual-Polarization\\Coherent  Receiver};

\node[block,right=0.3cm of rx] (adc){Analog-to-Digital\\ Conversion};

\node[dsp,above right=-0.4cm and 7.5cm of seq] (dsp1){Matched filtering~\cite{dorize18}};
\node[dsp,below=3mm of dsp1] (dsp2){Fiber detection};
\node[dsp,below=3mm of dsp2] (dsp3){Round-trip\\Jones-matrices extraction};
\node[dsp,below left=5mm and -17mm of dsp3] (dsp4a){$\widehat{\phi_k}(n)$ extraction~\cite{Guerrier20}};
\node[dsp,below=3mm of dsp4a] (dsp4b){$\widehat{\Delta\phi_k}(n)$ computation};
\node[dsp,below right=5mm and -19mm of dsp3] (dsp5a){Matrix product \& Eigen-\\value computation~\cite{Prato26}};
\node[dsp,below=3mm of dsp5a] (dsp5b){$\widehat{\Delta\beta_k}(n)$ Extraction};

\node[
draw,
dashed,
rounded corners,
fit=(dsp1)(dsp4b)(dsp5b),
inner sep=2mm,
label=above:{Digital Signal Processing}
](DSP){};


\draw[line] (laser)--(tx);
\draw[line,blue] (seq)--(tx);
\draw[line] (tx)--(circ);
\draw[line] (circ)--(fiber);
\draw[line] (circ)--(rx);
\draw[line]
(laser.south)
|- node[pos=.20,left]{LO}
(rx.west);
\draw[line,blue] (rx)--(adc);
\draw[line,blue] (adc.east)-- ++(0.3,0) |- (dsp1.west);
\draw[line] (dsp1) -- (dsp2);
\draw[line] (dsp2) -- (dsp3);
\draw[line] (dsp3.south) -- ++(0,-0.2) -| (dsp4a.north);
\draw[line] (dsp3.south) -- ++(0,-0.2) -| (dsp5a.north);
\draw[line] (dsp4a) -- (dsp4b);
\draw[line] (dsp5a) -- (dsp5b);
\end{tikzpicture}
\caption{On the left, a block diagram of a dual-polarization DAS system (blue (resp. black) arrows for electrical (resp. optical) signals) and on the right, the main digital-signal-processing steps. (I/Q: in-phase and quadrature, LO: Local Oscillator, $k$ is the spatial segment index and $n$ is the time index, $\widehat{\phi_k}(n)$ (resp. $\widehat{\Delta\phi_k}(n)$) is the estimated absolute (resp. differential) phase of segment $k$ at time $n$, $\widehat{\Delta\beta_k}(n)$ is the estimated local birefringence.)}
\label{fig:DASscheme}
\end{figure*}

DAS has evolved into a broad family of interrogation techniques that differ in their probing strategy, detection architecture, and the physical quantity extracted from the Rayleigh backscattered signal. Although these techniques rely on the same distributed scattering process, they exhibit distinct trade-offs in terms of spatial resolution, sensing bandwidth, sensitivity to strain, implementation complexity, and compatibility with deployed telecommunication networks. Hence, we start this section by reviewing the classifications of DAS techniques before discussing coexistence of DAS with data transmission.

\subsection{Receiver-based classification}
A first classification can be made based on the receiver architecture. Direct-detection (DD) systems recover only the intensity of the beating terms resulting from the Rayleigh-backscattered signal portions unless an interferometric process between the backscattered signal and a time-delayed version of it or a reference signal is performed before the detection~\cite{HartogChap6}. In contrast, coherent detection (CD) preserves both amplitude and phase of the optical field, enabling the extraction of distributed phase variations induced by external perturbations. While it offers important sensitivity gains compared to direct detection, laser coherence requirements are more stringent for CD-based systems. In its polarization-diversity version, coherent detection can also mitigate polarization fading that occurs when the state of polarization (SoP) of the backscattered signal and the SoP of the local oscillator are not aligned~\cite{Martins16,Yan17}. 

\subsection{Probing-based classification}
A second classification can be established according to the probing strategy. Table~\ref{table:1} summarizes a non-exhaustive list of representative DAS interrogation approaches and their relevance to deployed telecommunication networks. The reported results originate from different experimental configurations and are therefore not directly comparable (for instance in terms of noise floor). The table is intended to highlight the main trade-offs and demonstrated capabilities rather than establish a performance ranking. Conventional differential-phase-sensitive optical time-domain reflectometry ($\Delta\phi$-OTDR) relies on short optical pulses emitted from a highly-coherent laser for probing and estimates the local optical phase from the time-resolved Rayleigh backscattered signal. Localization of a strain event is then performed by computing the differential phase variations between two segments in the fiber separated by the so-called `gauge length'. Pulse-based interrogation naturally provides spatial localization through time-of-flight measurements. However, it suffers from signal-to-noise-ratio (SNR) limitations due to coherent fading, round-trip losses and laser phase noise. These impairments enforce a well-known trade-off between pulse duration, spatial resolution, sensitivity (i.e. the smallest detectable strain change observed by the propagating optical signal), sensing range (i.e. the maximum fiber length covered by DAS), and refresh rate. It also suffers from nonlinear effects at high peak power~\cite{Izumita94, Chiabaut26} and limited dynamic range due to phase unwrapping issues when the optical phase is locally disturbed by large strains. An alternative to the conventional $\Delta\phi$-OTDR that addresses the dynamic range issue is the chirped-pulse $\phi$-OTDR that moves the dynamic strain detection from a phase-based to a time-delay-based measurement~\cite{Pastor16}. 

Moreover, several other evolutions of the conventional $\Delta\phi$-OTDR were also proposed to enhance sensitivity. We cite two extensions: first, the use of multi-frequency pulses to mitigate coherent fading~\cite{Murray20,Wakisaka23}, and second, the use of linear-frequency-modulated (LFM) pulses, also referred to as time-gated OFDR (Optical frequency domain reflectometry) or even OPCR (Optical pulse compression reflectometry) and chirped-pulse DAS~\cite{Zou15, Wang17, Waagaard21, Ip22} in which the spatial resolution is determined by the sweeping range of the LFM pulses rather than by the pulse width. These LFM-pulse-based approaches help in enhancing the sensitivity and are very closely related to the schemes that will be introduced next. Alternatively, continuous-wave phase-encoded DAS probing techniques can be used to illuminate the fiber continuously (in contrast to pulsed-based probing) using deterministic probing sequences~\cite{dorize18,mompo19}, i.e. discrete modulated states such as phase-coded waveforms, or deterministic analog signals such as frequency sweeps also known as OFDR (Optical frequency domain reflectometry)~\cite{Shiloh17,Ryf26}. For both LFM-pulse-based DAS and continuous-wave DAS, spatial information is recovered through matched filtering~\cite{Ip22}, rather than through pulse propagation, enabling an important sensitivity enhancement while maintaining distributed measurements and avoiding the increase of the pulse peak power. Finally, dual-polarization probing was introduced in DAS~\cite{Guerrier20}, enabling an additional sensitivity enhancement and offering richer sensing capabilities as discussed in Section~\ref{sec:Birefringence}. We illustrate in the left part of Fig.~\ref{fig:DASscheme} the general scheme of a dual-polarization DAS probing system using I/Q-modulators.


\subsection{Coexistence of DAS and communication over a single unrepeatered link}
A single DAS interrogator can monitor tens or even more than a hundred kilometers of silica optical fibers at meter-scale spatial resolution and without the use of inline amplification~\cite{Waagaard21}. Two main deployment scenarios are typically studied over telecommunication optical networks. On the one hand, so-called `dark fibers' provide a relatively straightforward deployment of conventional DAS techniques and have already been employed in several field trials. On the other hand, lit fibers allow sensing over operational communication links, avoiding the need for dedicated fiber resources. However, this deployment scenario requires strict coexistence between sensing and communication signals, imposing constraints on optical power, spectral allocation, and network compatibility to preserve the quality of transmission (QoT)~\cite{Wang25}. In~\cite{Guerrier22}, in-band sensing and communication was achieved over a link of $80$~km with no impact on communication performance. A wavelength was dedicated to sensing and modulated with a constant-power ($<0$~dBm) continuous-wave DAS probe using polarization-multiplexed Golay-encoded sequences~\cite{dorize18} along with polarization-diverse coherent detection. The scheme corresponds to the one shown on the left of Fig.~\ref{fig:DASscheme}. The sensing wavelength was multiplexed within a C-band WDM transmission signal with a spacing of no more than $2$~GHz. Pulsed-based DAS interrogation struggles to achieve a similar performance due to the inherent nonlinear interference between the high-peak-power DAS signal and the neighboring communications signals~\cite{Wang25}.
 
Within this coexistence framework, the continuous-wave phase-encoded DAS occupies an intermediate position between conventional $\Delta\phi$-OTDR and OFDR. Like coherent $\Delta\phi$-OTDR, it relies on coherent detection of Rayleigh backscattering to monitor distributed phase variations. However, instead of using isolated pulses, the fiber is continuously illuminated with coded waveforms and the distributed response is reconstructed digitally through matched filtering. In this respect, the interrogation principle closely resembles OFDR, where the distributed impulse response of the fiber is also recovered from a continuously transmitted probing signal using coherent signal processing. Both approaches exploit long observation times, matched filtering, and digital reconstruction to maximize the collected backscattered energy. They differ primarily in the choice of the probing waveform and the domain in which the matched filtering is performed: optical or digital. Continuous-wave phase-encoded DAS interrogation was revisited by several research groups and was referred to by different names, such as spread-spectrum DAS, coded DAS, linear-frequency-modulated (or -swept) DAS, or even OFDR~\cite{Guerrier22,Guerrier20, dorize20, Mazur25, Ryf26}. OFDR estimates the impulse response of the fiber by probing the frequency response typically with a swept laser, whereas continuous-wave phase-encoded DAS estimates the same impulse response by probing the fiber with waveforms encoded with discrete symbols and applying correlation in the digital domain. Their key advantage over pulsed DAS techniques is the possibility of efficient integration into a shared spectral band with data-carrying WDM channels over lit fibers, with substantially reduced nonlinear interference.
 
\subsection{DAS deployment over an optical network}
Finally, DAS systems still need to overcome challenges imposed by the topologies of the optical communication networks. Two examples illustrate these challenges. First, DAS operation through predominantly unidirectional repeatered links is still impractical except over submarine optical links whose architecture enables several options for DAS integration without the filtering of the backscattered signals by the isolators in the amplification stages~\cite{Mazur25,Colares26}. Second, branching structures such as ROADMs, multiplexers and splitters bring additional problems for backscatter-based distributed sensing, as they severely attenuate sensing signals, generate branch ambiguities from multiple optical paths, and consequently complicate event localization and system calibration. Recent works have started addressing these issues in passive-optical-network (PON) scenarios using fast switching~\cite{Takahashi25} or wavelength and code orthogonality~\cite{Choudhury25,Wang26}.

  
\section{Fiber–Environment Coupling and Sensing Fidelity}\label{sec:Coupling}

\begin{figure}[ht]
    \centering
    \includegraphics[width=0.7\linewidth]{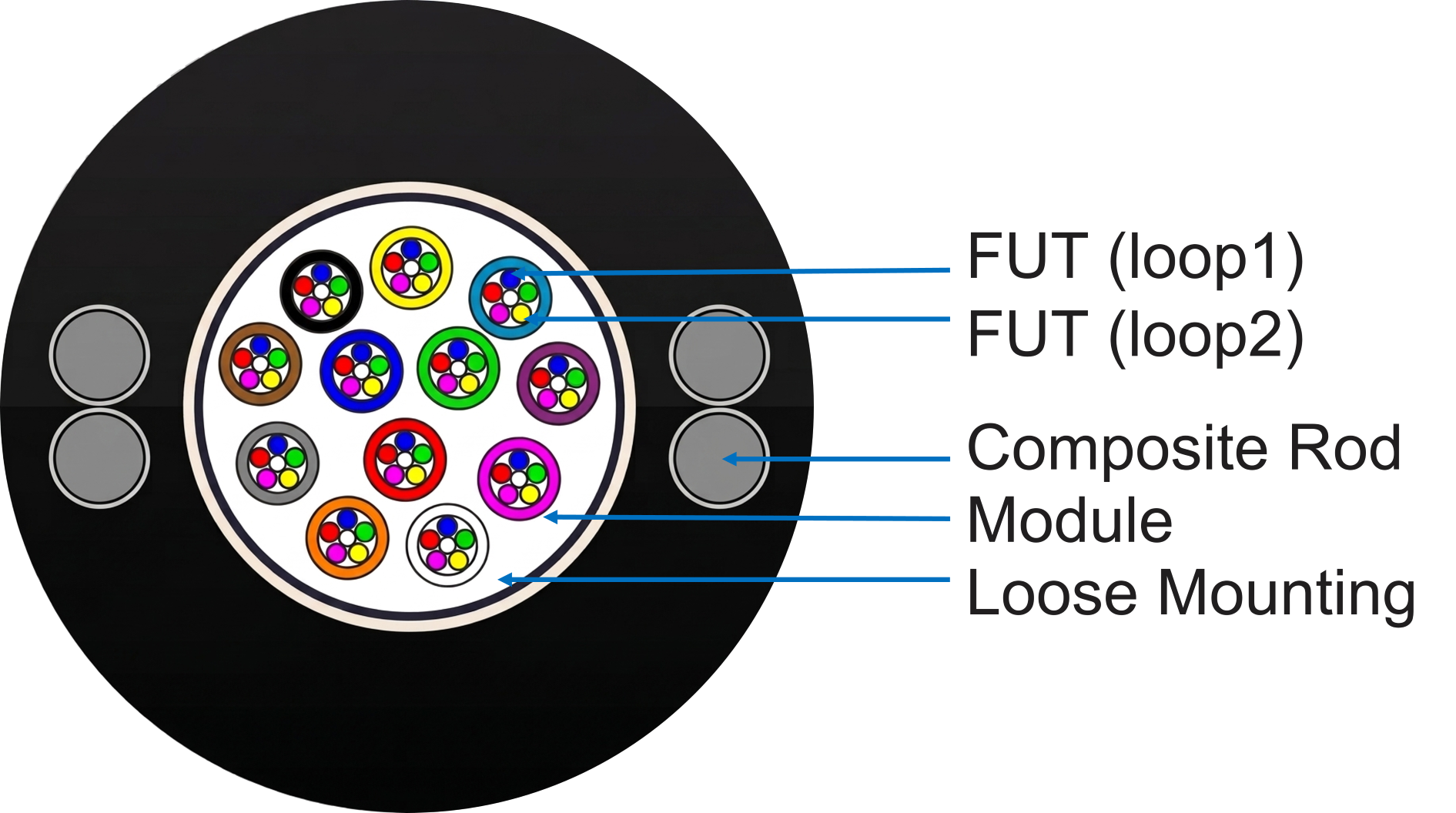}
    \caption{Cross section of the 72-fiber (SMF-G657) Prysmian L1091-14 cable.}
    \label{FUT}
\end{figure}

\begin{figure}[ht]
    \centering
    \includegraphics[width=1\linewidth]{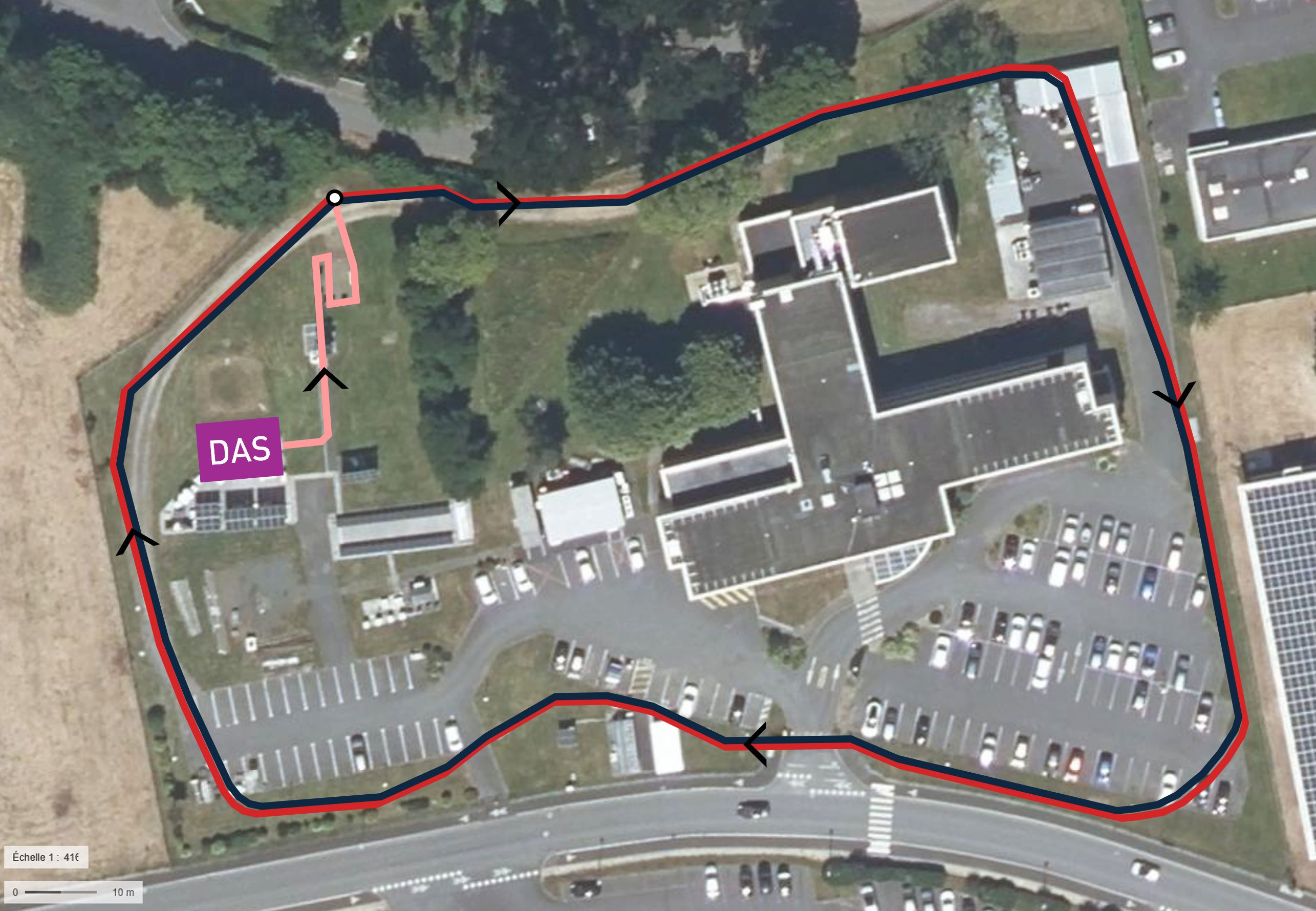}
    \caption{Orange Lab test field: FUT (Fiber Under Test) layout featuring a first section on the ground (in pink) followed by a first loop (in black), and a second loop (in red).}
    \label{Loops}
\end{figure}

A fundamental challenge in using deployed telecommunication fiber cables for quantitative distributed strain sensing is the variability of the coupling between the fiber cable and its surrounding media. Unlike dedicated sensing installations, telecommunication cables come in various configurations that are designed to limit the mechanical coupling to the surrounding media. They are also installed using diverse methods and paths, leading to non-uniform mechanical coupling along the cable. Recently, Orange Labs offered us the opportunity to perform DAS measurements on a deployed telecommunication cable in Lannion~\cite{Pruvost26}. The cable is a Prysmian L1091-14 containing 72 G.657 single-mode fibers with a total length of $1145~$m, as shown in Fig.~\ref{FUT}. It is installed as shown in Fig.~\ref{Loops}, with the first $64$~m (in pink) lying on the ground surface, followed by two loops of $540$~m that are underground (first loop in black and second loop in red). For the two loops, we used two optical fibers in the same module of an optical cable. The coupling of the cable with the surrounding media is not controlled and cannot be easily checked. During this field trial, surface waves were measured, and collocated sensors were used to verify the DAS measurements. More details on the experiments can be found in~\cite{Pruvost26}.

As a first conclusion, we noted that a quantitative strain-amplitude measurement is a significant challenge. The strain value in the surrounding medium can be different from the one measured in the glass fiber through DAS due to the impact of the coupling between the medium, the cable and the sensing fiber within the cable.
A promising metric to characterize the quality of the coupling between the optical fiber and the medium is the coherency metric introduced in~\cite{hudson2025b}. The coherency is a metric that compares neighboring DAS channels (i.e. fiber segments) to identify those that are observing the same strain variations. Its use is relevant when the observed mechanical perturbations have wavelengths $\lambda_{meca}$ that are greater than twice the gauge length, denoted $\Delta z$. In this case, the two neighboring DAS channels should measure the same strain perturbation. However, as we will see next, this metric cannot establish a quantitative measurement of fiber-cable-medium
coupling in complex environments such as urban ones. In the next paragraph, we explain how we adapt the coherency metric to our DAS measurements. 

Conventional DAS allows the extraction of differential phases $\widehat{\Delta\phi}_k(n)$, where $n$ is the slow-time index with a refresh rate of $6.1$~kHz and $k\in [1, K]$ is the DAS channel index. We apply the coherency metric to our field measurements using three gauge lengths: one equal to the native sampling resolution, $0.51$~m in our case, as well as $5.1$~m and $10.2$~m. We do not apply any processing to mitigate coherent fading and keep an equal spacing between the DAS channels. The only pre-processing step consists of detrending. $\widehat{\Delta\phi_k}(n)$ are zero-meaned over the slow-time dimension and are band-pass filtered with cut-off frequencies of $3$ and $15$~Hz to remove the slow variations, focusing on ambient surface waves induced by anthropogenic or natural sources. We recorded 16 five-second DAS measurements of ambient surface waves. Evaluating coherency using ambient acoustic noise follows a pragmatic approach developed in~\cite{hudson2025b} to qualitatively assess if a DAS channel is poorly or well coupled to its environment. The approach works the best when the energy of ambient surface waves is homogeneously distributed along the deployed cable, which would allow the isolation of intrinsic inter-segment coupling from any coherence induced by a dominant local event. Our results below will show that this condition is rarely met in complex urban environments.

For each DAS recording, the differential phases are cut into $W=8$ time windows of $T=6103$ samples (corresponding to $1$s) each with an overlap of $50\%$. 
For each time window, the differential phases are zero-meaned and normalized by their Euclidean norm (Eq.(1)). Normalization is required to assess relative coupling, as differential phases from successive channels may exhibit different amplitudes. Then, we compute the cross-correlation between successive channels $\widehat{\Delta\phi}_k(w,t)$ and $\widehat{\Delta\phi}_{k+1}(w,t)$ (Eq.(2)) where $w \in [1, W]$ is the time-window index and $t$ is a local time index in each window $w$. Finally, we keep the maximum absolute correlation value (Eq.(3)).

\begin{equation}
\Delta\phi^\prime_k(w,t) = \frac{\widehat{\Delta\phi}_k(w,t) - \frac{1}{T}\sum_{t=1}^{T} \widehat{\Delta\phi}_k(w,t)}{\sqrt{\sum_{t=1}^{T}\left(\widehat{\Delta\phi}_k(w,t) - \frac{1}{T}\sum_{t=1}^{T} \widehat{\Delta\phi}_k(w,t)\right)^2}}
\end{equation}\label{eq:CentRed}
\begin{equation}   
    r_k(w,\tau) = \sum_{t=\max(1,\,1-\tau)}^{\min(T,\,T-\tau)} \Delta\phi^\prime_k(w,t)\Delta\phi^\prime_{k+1}(w,t+\tau)
\end{equation}\label{eq:CrossCorr}
\begin{equation}
C_k(w) = \max_{\tau \in [-\tau_{\max},\, \tau_{\max}]} \left| r_k(w,\tau) \right| 
\end{equation}\label{eq:CohW}
where $\tau_{max}$ is considered for a surface wave with velocity $v_{\text{meca}}=300\text{m/s}$, hence $-3\text{ms} < \tau < 3\text{ms}$ with $\tau = 2\Delta z_{\mathrm{min}}/v_{\text{meca}}$ and $\Delta z_{\mathrm{min}}=0.51$~m. The absolute value operator is used to focus on the coupling strength and not its polarity. Hence, $C_k(w)$ only informs us if adjacent channels are reacting the same to ambient surface waves without considering any phase inversion. The maximization over the considered time-delay interval allows for small propagation-time differences between neighboring channels, but it may also increase the correlation between signals that are not necessarily generated by the same source. Consequently, the resulting coherency should be interpreted as a qualitative measure of similarity between neighboring channel responses, rather than as proof that both channels are observing the same physical excitation.

\begin{figure}[t]
    \centering
    \includegraphics[width=1\linewidth]{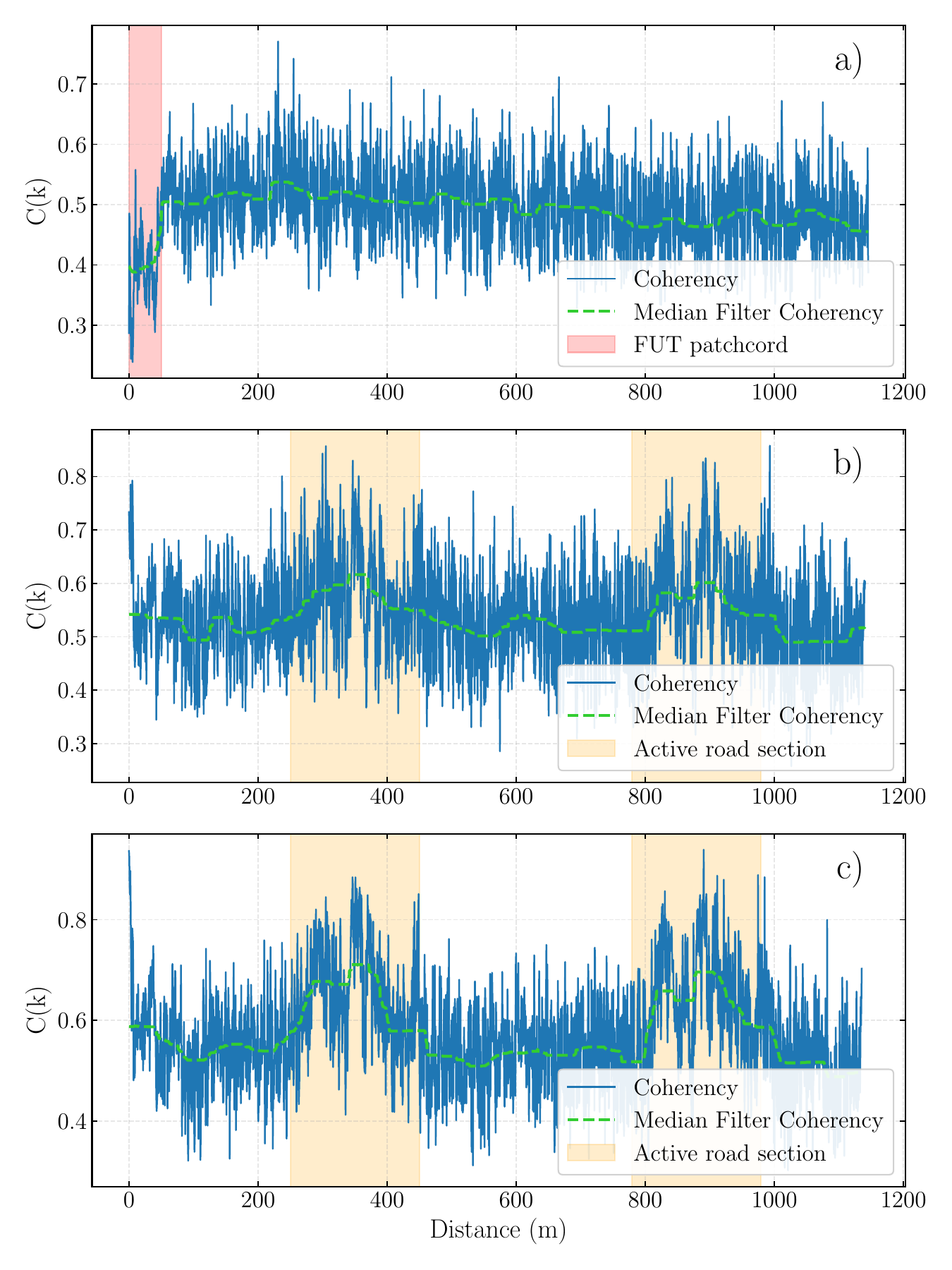}
    \caption{Coherency between adjacent channels with gauge lengths: a) $0.51$~m (native resolution), b) $5.1$~m and c) $10.2$~m.}
    \label{fig:Coherency_Lannion}
\end{figure}

For each DAS channel $k$, the computed coherency values are ranked and the $90\%$ of windows with the highest coherency values are retained to reduce the impact of noise and transitory acoustic events in the measured windows. The number of kept windows is denoted $n_\text{keep}$ and the formed set is denoted $W_\text{keep}$. 
From this subset, we calculate the mean coherency between adjacent channels as:
\begin{equation}
C(k) = \frac{1}{n_\text{keep}} \sum_{w \in W_\text{keep}} C_k(w)
\end{equation}
$C(k)$ is computed for each of the $16$ recordings and the stacked $C(k)$ curves are shown in Fig.~\ref{fig:Coherency_Lannion} in blue for the three gauge lengths respectively. The dashed green curve shows a median-filtered version of $C(k)$. First, when $C(k)$ is computed with a gauge length equal to the native resolution, we can clearly identify two average coherency levels: A $\sim60$~m-long section shows a much lower level than the rest of the link suggesting a lower mechanical coupling of that section with its surroundings and a potential higher sensitivity to high-frequency diffuse atmospheric noise. On-site observation of the link shows that the first $64$-m section lies on the ground surface or suspended in a technical cabinet before entering into the ground. 

\begin{figure}[t]
    \centering
    \includegraphics[width=1\linewidth]{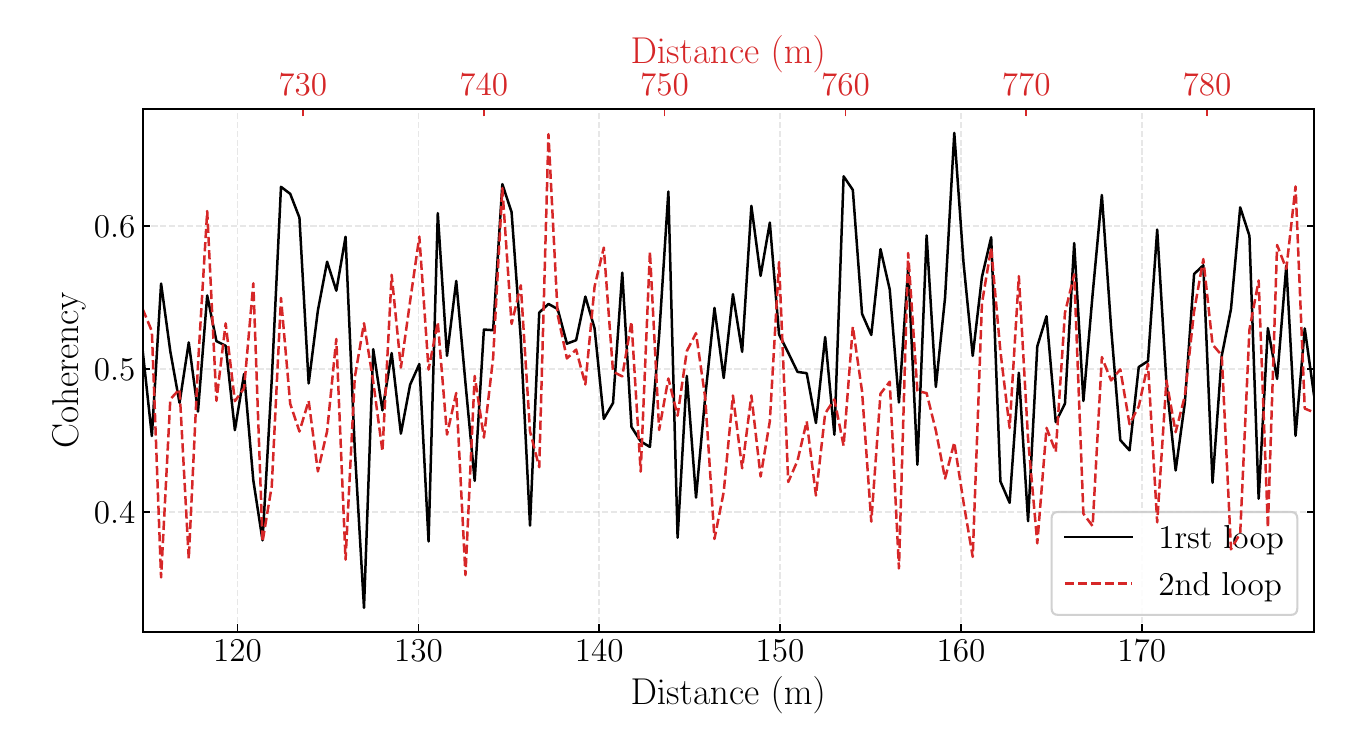}
    \caption{Zoomed-in view of the coherency profiles for a portion of the optical fiber common to both loops (gauge length of $0.51$~m).}
    \label{fig:Coherency_Zoom}
\end{figure}

Second, if we look more closely to the coherency levels of the underground sections of the link (Fig.~\ref{fig:Coherency_Lannion}(a) and Fig.~\ref{fig:Coherency_Zoom} in which we superpose the coherency levels of the two loops) and recall that these sections consist of two fiber loops installed in the same module of the same optical cable, we notice that the two fibers show different coherency levels with their surroundings even though they follow the same path within the same module. However, a different conclusion would be drawn when comparing the coherency levels of segments when differential phases are computed using larger gauge lengths of $5.1$~m and $10.2$~m respectively (Fig.~\ref{fig:Coherency_Lannion}(b,c)). For these values, the coherency level of the $60$-m fiber section on the ground surface increases relative to that computed at the native resolution, and the coherency levels of the two loops increasingly look alike. We explain these observations by a dependence of the coherency metric on the nature of the captured surface waves at each gauge length. Increasing the gauge length acts as a spatial low-pass filter that preserves the signatures of low-frequency surface waves in the differential phases. Given that low-frequency surface waves usually experience less attenuation in the ground than higher-frequency waves, we believe that the coherency levels obtained with larger gauge lengths are less affected by noise. Nevertheless, we cannot conclude from these measurements that the coherency levels are governed solely by the quality of the fiber-medium coupling. They also depend on the dominant surface waves in each DAS recording. For instance, we think that the high coherency levels in the yellow zones in Figs.~\ref{fig:Coherency_Lannion}(b) and (c) are due to the proximity of those fiber sections to a busy road; at these locations, we may have both well-coupled fiber segments to a homogeneous asphalt composition of the ground and dominant low-frequency surface waves. We may not isolate both effects. To this end, computing the coherency metric over much longer recordings may help in reducing noise.


Our observations are consistent with the findings in~\cite{Fairweather,Chedid} that reported variations in the DAS response of neighboring fibers within the same submarine telecommunications cable~\cite{Fairweather} or fibers with different coatings and deployment conditions in the same ground type~\cite{Chedid}, highlighting the influence of cable structure and fiber placement on DAS measurements. Whereas their study investigated controlled seismic and acoustic excitations, our results show that inter-channel coherency may provide a practical indicator of spatial variations in mechanical coupling along an operational terrestrial link. From these preliminary coherency evaluations on an installed telecommunication cable, we identify a significant challenge, but also an opportunity for further research toward a general calibration method for arbitrary deployed fibers. The calibration method could assess the coupling of each optical fiber with its surroundings and help determine the impact of different factors on DAS measurements, such as cable installation, the loose or tight mounting of fibers within the cable, coating conditions, cable topology, and the ground material surrounding the cable.

\section{Polarization Dynamics and Birefringence-Based Sensing}\label{sec:Birefringence}


Conventional DAS systems rely on common optical phase measurements, thereby neglecting polarization dynamics. However, real-world perturbations can also induce changes in the local fiber's birefringence. A recent Multiple-Input-Multiple-Output (MIMO) DAS system enabled the extraction of both phase and polarization parameters along the fiber~\cite{Guerrier20}; however, the main target was the mitigation of polarization fading effects. By further exploiting the estimated parameters for assessing the distributed linear birefringence strength, such DAS systems can detect and localize non-axisymmetric perturbations. Inspired by polarization-OTDR techniques mostly applied for static birefringence measurements~\cite{Palmieri13}, we recently investigated the use of polarization-diversity DAS and proposed a numerical model of polarization dynamics in silica fibers for DAS applications~\cite{Prato26}. In this section, we experimentally demonstrate additional static and dynamic strain sensing capabilities. Further details of the experimental setup can be found in~\cite{Prato26_2}. 

We note the round-trip matrix from the fiber start to segment $i$ and back $\mathbf{\widehat{H}}_i^{rt}$, for $i \in [1, N_{\mathrm{seg}}]$ with $N_{\mathrm{seg}}$ the total number of segments. In our setup, the sampling frequency at the DAS acquisition card is set to $400$~MHz, yielding a native sampling DAS channel length of $\delta z=0.255$~m. We send Golay codes of duration 0.163 ms (Golay order 12).  To mitigate coherent fading and therefore avoid large estimation errors, we can select the highest reflecting segments among groups of segments of a given size, resulting in a non-uniform gauge length. Hereafter, we will refer to the group size as the Low Resolution Factor (LRF). The selected subset of segments $J$ is of size $\lfloor N_{\mathrm{seg}}/\mathrm{LRF}\rfloor$ and the resulting average gauge length is $\overline{\Delta z}=\delta z\mathrm{LRF}$. For each consecutive pair of selected segments $(k, k')$ in $J$, we compute \cite{Galtarossa08, Feng18, Costa23}: 
\begin{equation}  
\mathbf{T}_{k'} = (\widehat{\mathbf{H}}_{k}^{rt})^{\dagger}\widehat{\mathbf{H}}_{k'}^{rt}\end{equation} 
Extracting the eigenvalues from this matrix product, we can compute an estimate of the effective retardance $\widehat{\Delta \beta_{\mathrm{eff}}L}$ \cite{Prato26_2}  (the phase shift between the two eigen-polarizations introduced by the propagation in the segment, due to the difference in refractive indices between the two modes), with $\widehat{\Delta \beta_{\mathrm{eff}}}$ the estimated effective birefringence magnitude in rad/m and $L$ the distance between segments $k$ and $k'$. The birefringence $\widehat{\Delta\beta_{\mathrm{eff}}}$ is an effective quantity that depends on the LRF. 

\begin{figure}
    \centering
    \includegraphics[width=1\linewidth]{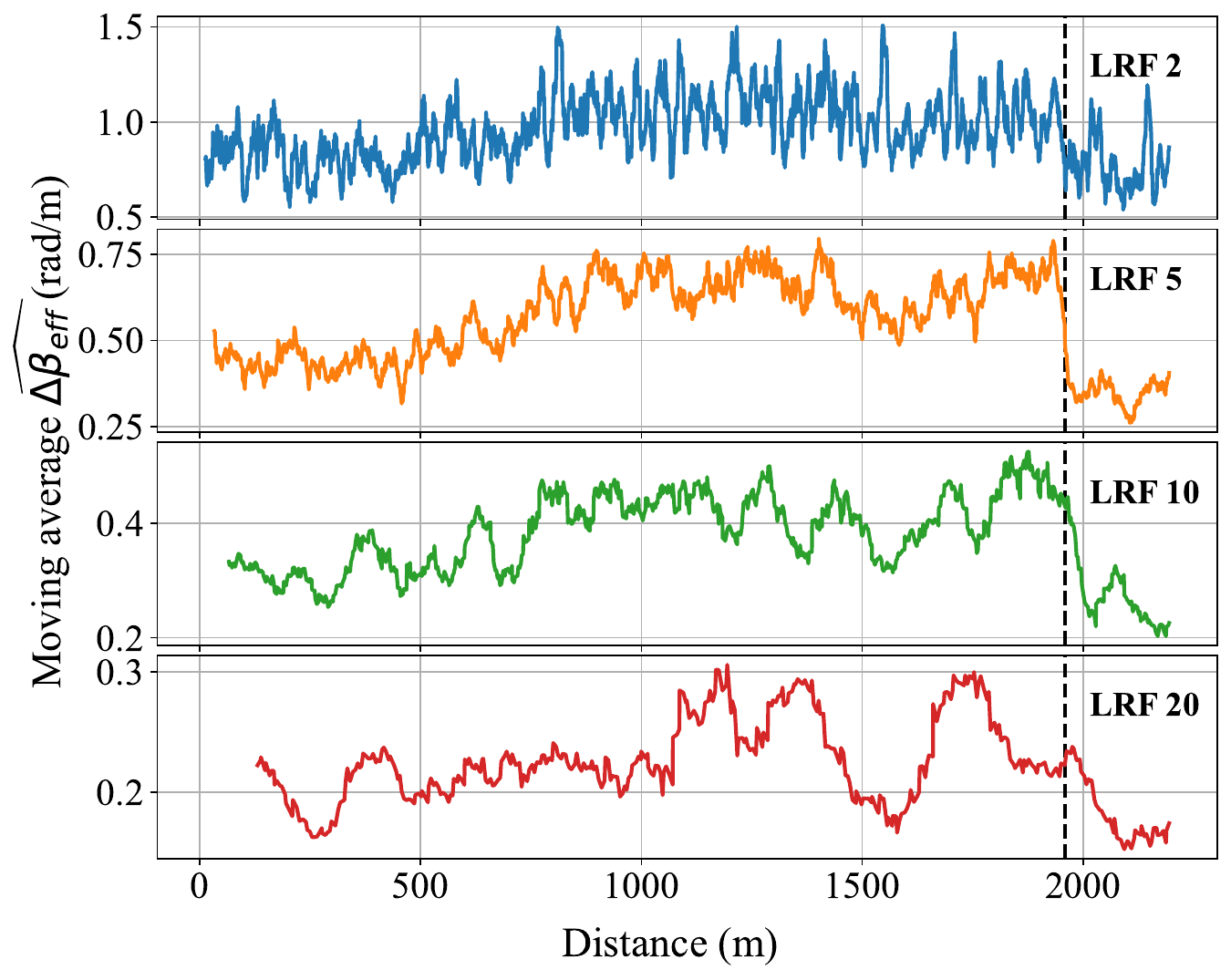}
    \caption{Effective birefringence magnitude $\widehat{\Delta \beta_{\mathrm{eff}}}$ versus distance for different LRFs and with a moving spatial average of $25$ segments. The transition between the two spools is marked by the red vertical line.}
    \label{fig:birefringence_vs_distance}
\end{figure}

When the local birefringence vector varies over length scales smaller than the distance between two selected segments, the estimated birefringence magnitude does not correspond to the local birefringence magnitude at a specific position. Instead, it results from the accumulation of the local birefringence vectors over the considered length. For illustration, we first performed measurements in static conditions over a $1958$~m fiber spool.\begin{figure*}
    \centering
    \includegraphics[width=1\linewidth]{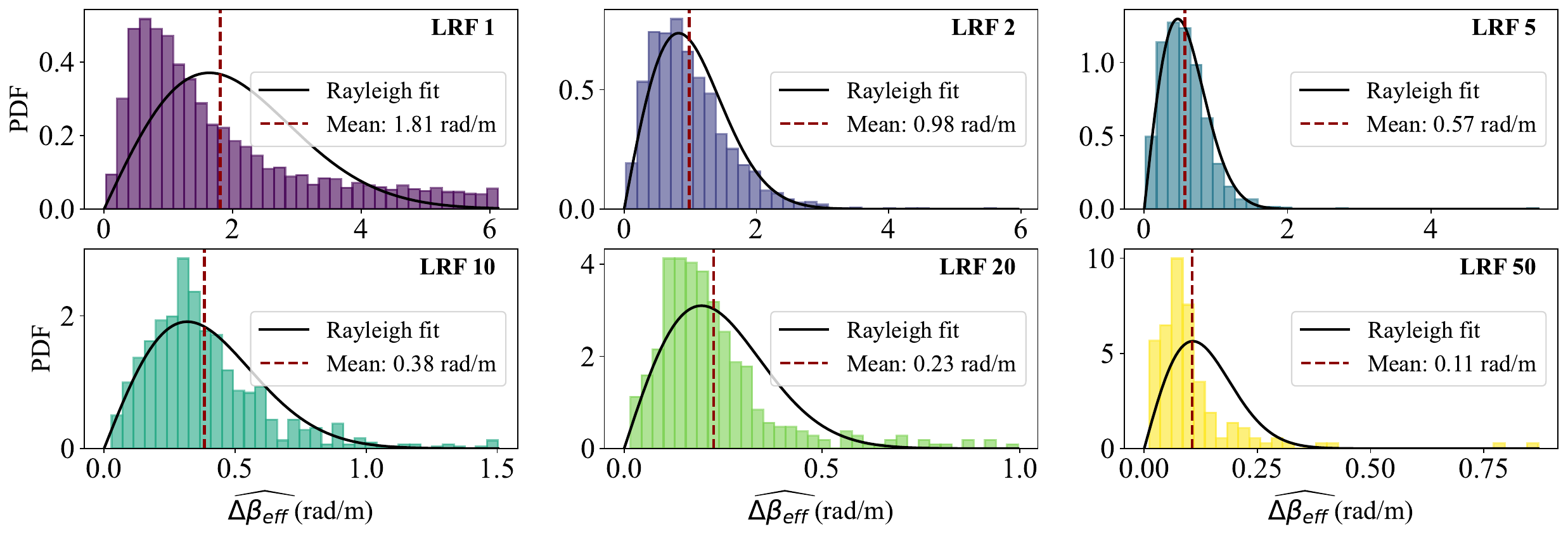}
    \caption{Histograms of estimated birefringence $\widehat{\Delta\beta_{\mathrm{eff}}}$ for different Low Resolution Factors (LRFs).}
    \label{fig:birefringence_histograms}
\end{figure*} Fig.~\ref{fig:birefringence_histograms} shows the distribution of the effective birefringence magnitude for different LRFs. When there is no selection, we observe a widespread distribution that does not follow the expected Rayleigh distribution~\cite{Wuilpart01}. This can be explained by the strong effect of coherent fading: the least reflective segments lead to large errors in the estimated matrices, which are not reliable. For LRFs of 2 and 5, we see that the distributions better show the expected behavior. However, for higher LRFs, the distributions tend to deviate from a Rayleigh distribution. We observe that as the LRF (and consequently the gauge length) increases, the mean value of the estimated birefringence decreases. This effect has been observed in~\cite{Chen21} and might be explained by the fiber being spun. Indeed, the propagation over the gauge length can be viewed as the composition of successive polarization rotations on the Poincaré sphere. The equivalent rotation angle is always smaller than or equal to the sum of the individual rotation angles, with equality only when all local birefringence vectors are aligned. Consequently, changes in birefringence orientation reduce the magnitude of the effective birefringence relative to the true local birefringence. Therefore, the choice of the LRF is a trade-off between coherent fading mitigation and accurate birefringence estimation. Since the estimated birefringence depends on the spatial resolution, relative trends are more informative than absolute values. Hence, we performed a measurement over the $1958$~m fiber spool followed by a $235$~m fiber spool. Fig.~\ref{fig:birefringence_vs_distance} shows the birefringence estimation for different LRFs as a function of distance in the fiber. A moving spatial average with a window length of $25$ was applied to smooth the traces and highlight the main trends. On the traces, we can observe the transition from the first fiber spool to the second, especially for LRFs $5$ and $10$. For an LRF of $2$, the trace is still quite noisy, making the transition less clear, whereas for an LRF of $20$, the averaging becomes too large to see a sharp transition. Thus, the choice of the LRF needs to be tailored depending on the application. 

Finally, to assess the ability to detect dynamic events through birefringence estimation, we placed a fiber squeezer ($\sim1$~cm of fiber squeezed between two plates in a General Photonics PolaRITE II device) between the $1958$~m spool and $235$~m spool. This device compresses the fiber between two parallel plates to induce birefringence changes through a transverse force. Indeed, birefringence changes mainly occur from anisotropic strains that break the cylindrical symmetry. 
\begin{figure}
    \centering
    \includegraphics[width=1\linewidth]{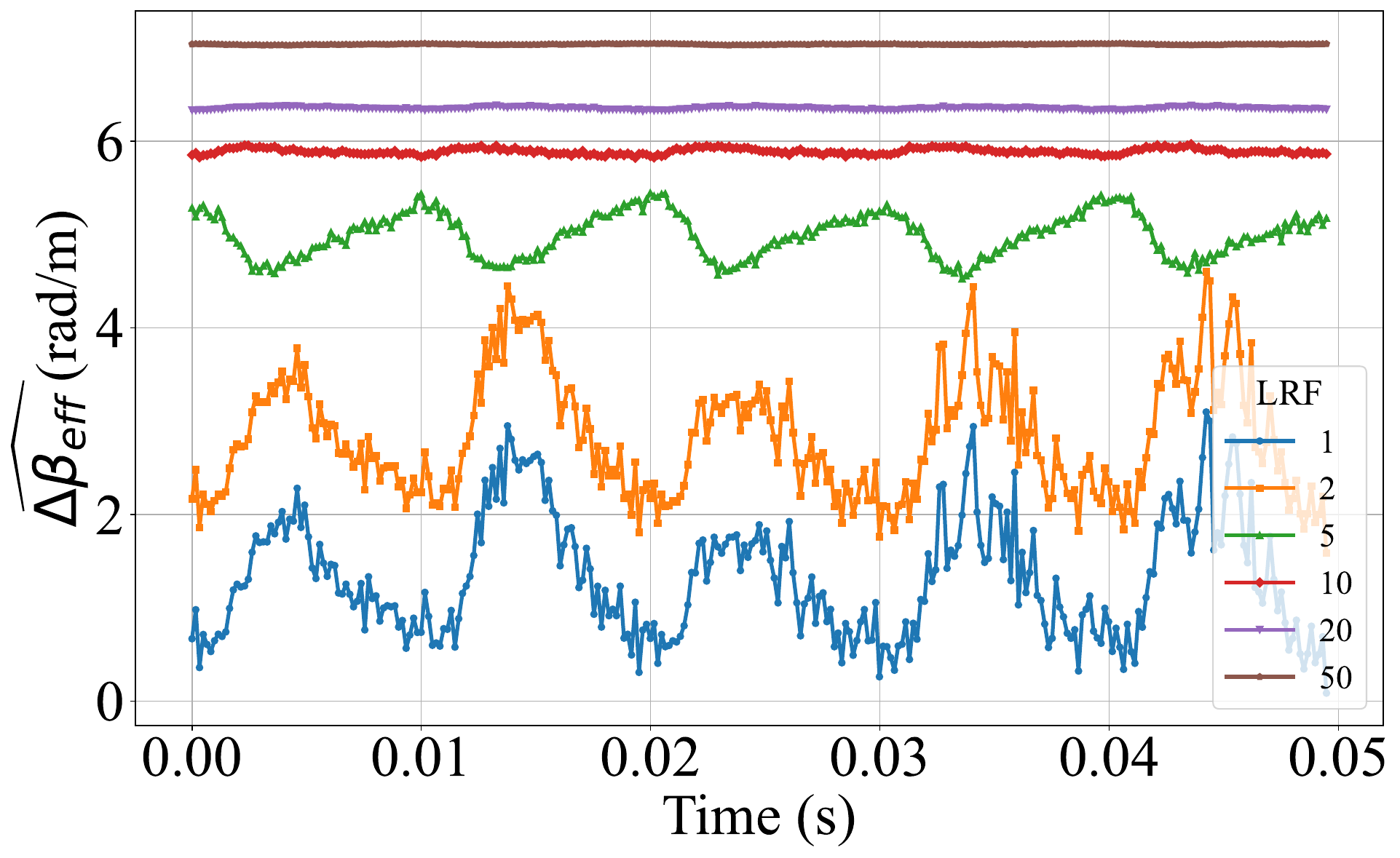}
    \caption{Effective birefringence estimation at event location versus time for different LRFs, with added offsets to separate the traces.}
    \label{fig:transverse_event}
\end{figure}
Fig.~\ref{fig:transverse_event} shows the effective birefringence magnitude estimated at the event location for different LRFs versus time, with added offsets for better visualization. 
For LRFs of $1$ and $2$, the estimated effective birefringence exhibits a similar temporal evolution, whereas the trace for $\mathrm{LRF}=5$ shows temporal variations of opposite sign. This difference may arise from the vectorial nature of birefringence accumulation over the measurement segment. Here, the two traces for LRFs $1$ and $2$ are identical because the two selected segments might be consecutive at a particular fiber location. To explain the birefringence evolution versus LRFs, we recall that the squeezer acts on a localized $1$~cm fiber section, which is shorter than the segment length for all considered LRFs. Consequently, the estimated effective birefringence reflects the combined contribution of the squeezer-induced birefringence and the intrinsic birefringence of the surrounding fiber sections, whose relative orientations may differ. Depending on these orientations and on the segment boundaries selected for each LRF, the resulting effective birefringence can exhibit different temporal variations. As the LRF increases, the contribution of the localized perturbation is increasingly diluted by the surrounding fiber, making the event signature almost unnoticeable at higher LRFs. 


\section{Phase Noise Limitation}\label{sec:PhaseNoise}

Laser phase noise is a fundamental limitation in coherent DAS systems~\cite{Loayssa22}, especially when applied over long fiber spans or when phase-encoded DAS or chirped-DAS interrogation along with matched filtering is used~\cite{Ip22}. Phase noise directly impacts the accuracy of channel estimation and the reliability of extracted phase and polarization parameters. Understanding and mitigating phase noise is therefore essential for next-generation DAS systems targeting long-range sensing over deployed telecom fibers with telecom-grade optical components. In this context, we present a comparative numerical study of coded sequences to assess their resilience to phase noise when using a MIMO DAS system.

\subsection{Studied probing sequences}
For this study, we select several sequence families employed in phase-encoded DAS systems: Golay complementary pairs~\cite{dorize18}, Legendre sequences~\cite{mompo19}, maximum length sequences (m-sequences)~\cite{Martins16}, and Constant Amplitude
Zero Auto-Correlation (CAZAC) sequences~\cite{dorize20}. Within the CAZAC family, we consider two variants: the Perfect-Square Minimum-Phase (PS-MP) sequence and Zadoff–Chu sequences~\cite{chu1972, andrews22}. 

Periodic autocorrelation is a key performance indicator for sequences employed in DAS systems. Among the selected families, Golay and CAZAC sequences provide perfect periodic autocorrelation zones, enabling ideal channel estimation, whereas Legendre and m-sequences exhibit non-perfect periodic autocorrelation (they tend toward perfect autocorrelation as their length tends to infinity).

Beyond autocorrelation, the probing sequences differ in their frequency diversity, here understood qualitatively as how broadly their energy is distributed across frequency.  A broad distribution provides greater frequency diversity, whereas energy concentrated over a narrower frequency range corresponds to lower diversity. The three Zadoff–Chu (ZC) CAZAC sequences considered here exhibit maximum, intermediate, and minimum frequency diversity (controlled by the choice of their root index~\cite{Freire26}), respectively, as illustrated by their spectrograms~\cite{Freire26}. The PS-MP CAZAC sequence presents a narrow frequency diversity, while the Golay, Legendre and m-sequences have approximately uniform spectral distributions and therefore broad frequency diversity. This distinction is relevant to phase-noise resilience, as discussed below. The ZC and PS-MP CAZAC sequences are also related to LFM waveforms through their quadratic phase profiles, whose phase increments evolve linearly with the symbol index. For ZC sequences, the root index determines the phase progression and thus the associated discrete chirp slope, subject to the sequence length and discrete-time representation. PS-MP CAZAC sequences likewise exhibit a chirped phase structure, although their construction differs from that of ZC sequences. Despite this connection, the discrete sequences and a continuous LFM pulse are not strictly interchangeable: their spectral properties and correlation characteristics depend on the specific waveform construction.

Several practical distinctions between the sequence families should be noted. CAZAC sequences are composed of many discrete unit-norm complex-valued symbols, which increases their hardware implementation complexity because a very accurate I/Q electro-optic modulator would be needed to generate them. Among them, only the PS-MP CAZAC sequence can be directly mapped onto Phase Shift Keying (PSK) modulation. All remaining families are binary. Furthermore, only Golay sequences provide two distinct mutually orthogonal pairs, one per polarization; for all other families, orthogonality between polarizations is obtained by cyclically shifting the sequence transmitted on the second polarization by half the sequence length~\cite{dorize20}. Finally, the sequences cannot be constructed for arbitrary lengths if their autocorrelation properties are to be preserved, which constrains the code lengths available to each family.

\subsection{Initial back-to-back study with time delay}
Before evaluating the sequences within the DAS simulation framework, we investigate the effect of phase noise on the sequence correlation properties. For this purpose, the correlation is computed between the original undisturbed sequence, $s[n]$, and its phase-noise-impaired counterpart,
 
\begin{equation}\label{eq:pn_affected_sequence}
    s'[n] = s[n] e^{j \phi_{TX}[n]} e^{-j \phi_{LO}[n]} = s[n] e^{j (\phi[n] - \phi[n-d]}),
\end{equation}
 
\noindent where $\phi_\mathrm{TX}[n]$ and $\phi_\mathrm{LO}[n]$ denote the transmitter (TX) and local oscillator (LO) phase noise, respectively. Since both originate from the same laser and follow a random-walk process $\phi[n]$, they differ only by the round-trip delay $d$. Although this scenario corresponds to a back-to-back configuration without fiber propagation, the TX--LO delay can be related to the round-trip delay of the different fiber segments: shorter delays represent segments closer to the interrogator, while longer delays correspond to more distant points along the fiber. Two delays of 1 and 1000~symbols are considered, both much shorter than the sequence lengths, which are approximately 16384~symbols, with slight differences between sequences.
 
\begin{figure}[t]
    \centering
    \includegraphics[width=\linewidth]{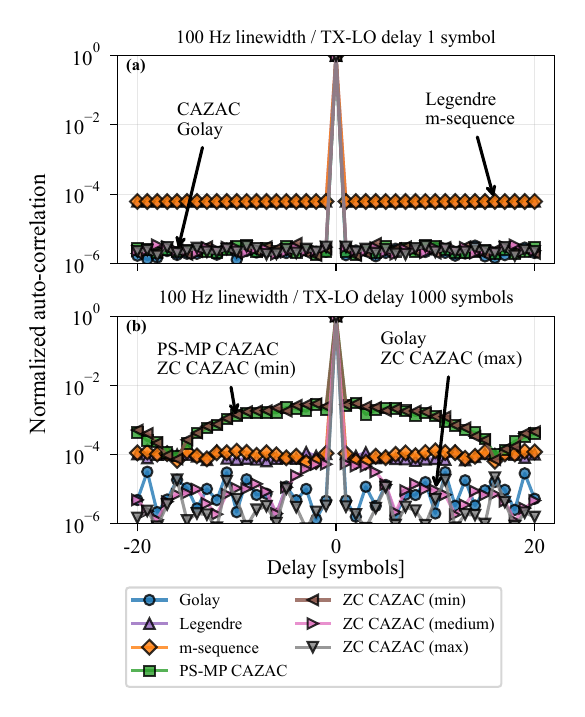}
    \caption{Correlation of each of the considered coded sequences with its phase-noise-affected counterpart, zoomed around the correlation peak, for a 100~Hz laser linewidth and a TX--LO delay of (a) 1~symbol and (b) 1000~symbols. In (b), the intermediate-diversity ZC CAZAC sequence lies between the two labelled groups.}
    \label{fig:b2b-pn-correlation}
\end{figure}
 
Fig.~\ref{fig:b2b-pn-correlation}(a) shows the normalized correlation for a $100$~Hz FWHM laser linewidth and a one-symbol TX--LO delay. Legendre and m-sequences exhibit sidelobes at approximately $6\times10^{-5}$, a level set by their non-perfect autocorrelation and already present in the absence of phase noise. Golay, PS-MP CAZAC and ZC CAZAC sequences, in contrast, lose their perfect autocorrelation and develop sidelobes of the order of $10^{-6}$, still more than one order of magnitude below the sidelobes of the sequences with non-perfect autocorrelation.
 
For the 1000-symbol delay in Fig.~\ref{fig:b2b-pn-correlation}(b), the degradation becomes considerably more pronounced and a clear ordering emerges. Legendre and m-sequences remain close to their phase-noise-free level, while PS-MP CAZAC and the minimum-diversity ZC CAZAC sequence exhibit the highest sidelobes, exceeding $10^{-3}$. Golay and the maximum-diversity ZC CAZAC sequence show the lowest levels, with the intermediate-diversity ZC CAZAC sequence lying in between. This ordering follows the frequency diversity of the sequences. These results demonstrate that a higher frequency diversity provides robustness against phase-noise-induced degradation of the channel estimation.

\subsection{Numerical study over a sensing fiber}
To evaluate the performance of these sequences on the MIMO DAS framework, simulations are performed at a symbol rate of $50$~Mbaud corresponding to $2$~m segments, over a $2$~km fiber, and with variable code lengths. Each configuration is repeated for $10$~independent Rayleigh-distributed fiber profiles, and the reported results correspond to the average across all realizations. Differential phases are extracted following the DSP steps shown in the right part of Fig.~\ref{fig:DASscheme}. Phase errors are evaluated with respect to the ground-truth values available in simulations, after selecting one out of every ten scatterers based on the reflectivity, resulting in an effective gauge length of approximately $20$~m.

\begin{figure*}[!t]
    \centering
    \includegraphics[width=1.0\linewidth]{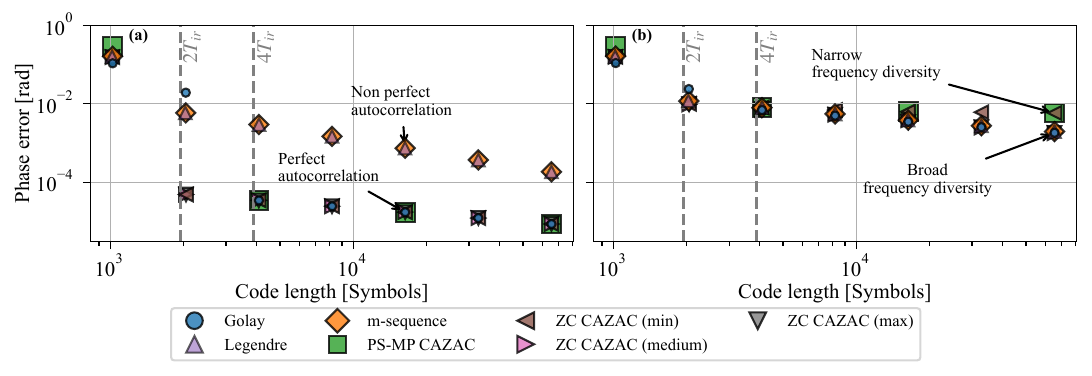}
    \caption{Phase estimation error versus code length for the considered sequences, simulated over a 2~km fiber (a) without phase noise and (b) with phase noise corresponding to a 10 Hz laser linewidth. Vertical dashed lines mark twice and four times the channel spreading time of the fiber, $2T_{ir}$ and $4T_{ir}$.}
    \label{fig:sequences_phase_noise}
\end{figure*}

Fig.~\ref{fig:sequences_phase_noise}(a) shows the phase error for each sequence as a function of code length when phase noise is excluded from the simulation framework. For the shortest code lengths, all sequences exhibit large errors, since the code duration is shorter than twice the channel spreading time of the fiber, $2T_{ir}$, which is the minimum required for full Jones matrix estimation of the round-trip propagation. As the code length increases, the errors decrease. CAZAC sequences are the first to do so, owing to their perfect autocorrelation properties; note that all CAZAC variants yield overlapping results in this scenario, and that PS-MP CAZAC sequences exist only for perfect-square lengths, restricting them to powers of four on the simulated grid, which explains the missing points for this sequence. Golay codes also possess perfect autocorrelation; however, they require code lengths of at least $4T_{ir}$ in order to avoid aliasing with their non-zero autocorrelation zone, beyond which their performance coincides with that of the CAZAC sequences. The remaining sequences, Legendre and m-sequences, whose results also overlap, show errors that decrease with code length at a slower rate, as a consequence of their non-perfect autocorrelation properties.

Fig.~\ref{fig:sequences_phase_noise}(b) shows the corresponding results when phase noise from an ultra-narrow-linewidth laser of $10$~Hz FWHM linewidth is included. Two effects are observed. First, the error reduction with code length slows down considerably; as a consequence, the advantage of perfect autocorrelation largely vanishes, and all sequence families converge towards a similar error level. Second, a distinct behaviour emerges depending on the frequency diversity of each sequence: the PS-MP CAZAC and minimum-diversity Zadoff–Chu sequences, both with narrow frequency diversity, reach an error approximately three times as high as that of the sequences whose energy is broadly distributed across the spectrum ($5.7 \times 10^{-3}$ versus $1.8 \times 10^{-3}$~rad at the largest code length considered). This behavior is consistent with the back-to-back correlation results of Fig.~\ref{fig:b2b-pn-correlation}(b) and may be attributed to an averaging effect: spreading the sequence energy over a wider frequency range averages the phase-noise-induced distortion across the spectrum, benefiting the estimation as was observed for Golay codes in~\cite{dorize20}. These results demonstrate that autocorrelation quality governs performance in the additive-noise-limited regime, whereas frequency diversity becomes a decisive sequence property once phase noise dominates. 

\subsection{On laser phase noise compensation}
Even when the sequence most resilient to phase noise is selected, performance degradation cannot be avoided. In such cases, phase noise compensation becomes necessary. Most existing compensation techniques, however, entail an increase in hardware complexity: recent works have focused on enhancing the phase stability of the laser source~\cite{Fontaine25}, while others measure the phase noise with an auxiliary coherent receiver and subsequently subtract it from the main sensing signal~\cite{Pineiro23}. Although effective, these approaches introduce additional hardware and cost. This motivates the investigation of phase noise estimation and compensation based purely on digital signal processing, requiring no additional hardware.

We address phase noise compensation through a joint scheme combining a Kalman filter~\cite{welch1995} for laser phase-noise estimation with least-mean-squares (LMS) MIMO channel estimation. For channel estimation, this scheme replaces the first three DSP operations shown in Fig.~\ref{fig:DASscheme}: matched filtering, fiber detection, and round-trip matrix extraction. Matched filtering may nevertheless be retained separately to identify the beginning and the end of the sensed fiber, thereby providing synchronization. The Kalman filter and LMS estimator operate jointly on a sample-by-sample basis. For each received sample, the Kalman filter first predicts the current laser phase noise, and the predicted phase is used to de-rotate the sample before the LMS estimator updates the round-trip channel matrix. The residual phase deviation is then used to update the Kalman state before processing the next sample. Thus, the two estimators jointly recover the phase noise and the channel response, rather than estimating the channel by conventional matched filtering.

The laser phase noise is modeled as a Wiener process. Accordingly, the phase-increment process-noise variance in the Kalman filter is set to $Q=2\pi\Delta\nu T$, where $\Delta\nu$ is the laser linewidth and $T$ is the sampling interval along the fast-time axis. The measurement-noise variance $R$ is selected by minimizing the root-mean-square (RMS) phase error relative to the ground-truth phase, which is possible in simulations but would require an independent reference measurement in a field experiment. 
The filter is applied sequentially along the fast-time sample axis, which maps to position along the fiber, rather than across successive interrogations at a fixed position. Since the transmitter and local oscillator fields originate from the same laser and are separated by the fiber round-trip delay, their phase-noise difference depends on the delay associated with the position being interrogated. A single phase noise process suffices to describe both transmitter and local-oscillator contributions, provided the delay is known; the Kalman filter therefore estimates only this single process. 

To evaluate this approach, we perform simulations over a range of laser linewidths, comparing the cases with and without phase noise compensation. For the same 2~km fiber as before, Golay complementary pairs of 4096~symbols are transmitted, and the results are averaged over 5 independent Rayleigh-distributed fiber realizations. The simulations are performed under static conditions, i.e., without applied strain events, so that any residual variation in the estimated channel originates from phase noise alone. The LMS step size is set to 0.1, chosen as a compromise: since both estimators operate jointly, a faster adaptation would allow the LMS to absorb part of the phase noise into the channel estimate, degrading the recovered sensing phase, while a slower one would limit the tracking of channel variations. Its optimization, and the evaluation under dynamic conditions, are left for future work. 

\begin{figure}[t]
    \centering
    \includegraphics[width=1.0\linewidth]{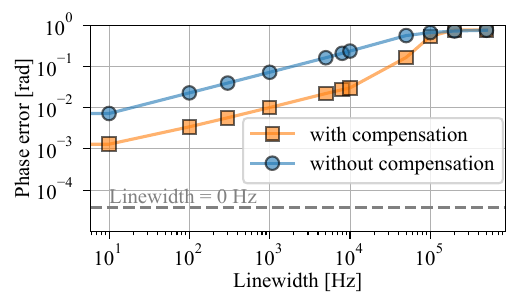}
    \caption{Phase error as a function of laser linewidth using Golay complementary pairs of 4096~symbols over a 2~km fiber. In the compensated case, the phase noise is estimated by a Kalman filter operating jointly with LMS MIMO channel estimation. The dashed line indicates the phase-noise-free baseline.}
    \label{fig:kalman}
\end{figure}

Fig.~\ref{fig:kalman} shows the resulting phase error as a function of laser linewidth values between $10$~Hz and $100$~kHz, together with the phase-noise-free baseline (dashed line). The per-sample tracking of the Kalman filter is beneficial over the entire linewidth range, with the compensated error remaining approximately a factor of five below the uncompensated case. The residual gap to the phase-noise-free baseline reflects the estimation noise of the filter, which prevents perfect phase recovery even at the lowest linewidths. Above approximately 10~kHz, the phase evolves too rapidly for the filter to track accurately, and the benefit progressively narrows, until both cases converge, at linewidths beyond $\sim$100~kHz. 

This preliminary numerical analysis quantified phase-estimation error in phase-encoded MIMO DAS probing for a $2$~km fiber under static conditions and for a range of laser linewidth values. It showed that a Kalman-filter-based phase estimation and compensation can reduce the error over a large range of linewidths. However, phase-estimation error alone does not fully characterize sensing performance: the trade-off between phase-noise suppression, residual estimation error, and the minimum detectable strain remains to be quantified. Further work should therefore investigate this trade-off through experimental measurements under controlled dynamic strain conditions, over longer fiber reaches, and compare Kalman filtering with alternative compensation strategies. 

\section{Conclusion}\label{sec:Conclusion}


Distributed fiber sensing over deployed networks has the potential to transform these networks into pervasive monitoring platforms. 
In this paper, we shed light on major challenges facing the deployment of DAS over telecommunication networks. After a review of the main interrogation approaches and an emphasis on the importance of using constant-amplitude continuous DAS probing for in-band coexistence with data transmission in the same fiber, we focused specifically on three challenges: the dependence of DAS measurements on the cable deployment conditions, the extraction of birefringence measurements from MIMO DAS and the impact of laser phase noise on continuous-wave phase-encoded DAS. We briefly describe a few opportunities to address these challenges. For the first, the assessment of the coupling between the sensed fiber and its surrounding medium could be used to calibrate the extracted strain, offering a complementary tool to propagation delays or frequency-wavenumber ($f-k$) metrics. For the second, while we experimentally demonstrated that the proposed method enables extraction of static and dynamic birefringence-related information using a polarization-diversity DAS, a validation on a deployed link is still needed. Besides, as the event characterization depends on the chosen spatial resolution, multi-resolution detection approaches can help better understand the detected strains. For the third, we observed, through numerical simulations, that autocorrelation and frequency diversity properties must be considered in sequence selection for phase-encoded MIMO DAS systems. Moreover, digital signal processing-based phase-noise compensation could enable the use of lasers with larger linewidth, and therefore lower cost, by relaxing the phase noise requirements for a given target performance. The scheme illustrated in this paper is only one possible realization and its performance needs to be evaluated under field conditions. The residual estimation-noise floor and high-linewidth tracking limit leave room for improvement through other estimation and compensation approaches. 

Future research is expected to focus on the design of new network-compatible sensing architectures, further integration of sensing and communication, 
and management of the data volumes from large-scale sensing over operational networks.

\section*{Acknowledgments}
Our research works received funding from ANR (Agence Nationale de la Recherche) within the SAFER project (grant ANR-24-CE42-3114), from AID (Agence Innovation Défense) within the FIBROUS project, and from IMT (Institut Mines-Télécom) within the FRAME-XG program.

\vfill

\end{document}